\documentclass[12pt]{iopart}
\usepackage{graphicx}
\usepackage{epstopdf}
\usepackage{amsfonts}

\usepackage[linesnumbered,boxed,ruled,algonl]{algorithm2e}

\graphicspath{{img/}{.}{fig/}}

\begin{document}
\paper[Obtaining Communities with a Fitness Growth Process]{Obtaining Communities with a Fitness Growth Process}

\author{Mariano G. Beir\'o$^{1}$, Jorge R. Busch$^{1}$, Sebastian P.
Grynberg$^{1}$ and J. Ignacio Alvarez-Hamelin$^{1,2}$}

\address{$^1$ Facultad de Ingenier\'{\i}a,
         Universidad de Buenos Aires, Paseo Col\'on 850,\\
         ~\hspace{1.3mm}C1063ACV~Buenos Aires, Argentina}
\address{$^2$ INTECIN (CONICET--U.B.A.)}
\ead{mbeiro@fi.uba.ar}

\begin{abstract} 

The study of community structure has been a hot topic of research over the last years. But, while successfully applied in several areas, the concept lacks of a general and precise notion.
Facts like the hierarchical structure and heterogeneity of complex networks make it difficult to unify the idea of community and its evaluation. The global functional known as {\em modularity} is probably the most used technique in this area. Nevertheless, its limits have been deeply studied. Local techniques as the ones by Lancichinetti {\em et al.} and Palla {\em et al.} arose as an answer to the resolution limit and degeneracies that modularity has.

Here we start from the algorithm by Lancichinetti {\em et al.} and propose a unique {\em growth process} for a fitness function that, while being local, finds a community partition that covers the whole network, updating the scale parameter dynamically. We test the quality of our results by using a set of benchmarks of heterogeneous graphs. We discuss alternative measures for evaluating the community structure and, in the light of them, infer possible explanations for the better performance of local methods compared to global ones in these cases.

\end{abstract} 
\maketitle 

\newtheorem{theorem}{Theorem}
\newtheorem{corollary}{Corollary}
\newtheorem{lemma}{Lemma}
\newtheorem{definition}{Definition}
\newtheorem{references}{References}
\newtheorem{questions}{Questions}
\newcommand{\one}{\mathbf 1}
\newcommand{\prob}{\mathbf P}
\newcommand{\esp}{\mathbf E}
\newcommand{\cov}{\mathbf cov}
\newcommand{\sdi}{\bigtriangleup}


\section{Introduction}\label{sec:intro}

In the last years community detection became one of the top research topics in the area of Complex Networks. Due in part to the explosion of social networking, but also to its application in diverse areas as ecology and computational biology, an interest arose in defining, detecting, evaluating and comparing community structures. For a thorough -yet not exhaustive- reference of its applications see the survey by~\cite{Santo201075}.

The early research by Newman departed from the use of {\em betweenness} to divide the network into modules~\cite{girvan_2002}, and the definition of {\em modularity} to evaluate communities~\cite{newman:faecsin}. Then he proposed using the modularity as a functional to be maximized~\cite{newman:macsin}. Different optimization techniques were developed, of which we recall the algorithm by Guimer\`a based on simulated annealing~\cite{guimera} for its good results, and the Louvain algorithm~\cite{blondel:fuociln} for its fast convergence within large networks.

Later, the works by~\cite{clauset2010} and~\cite{fortunato:rlicd} questioned the global optimization methods based on modularity, for being prone to resolution limits and extreme degeneracies. Local techniques were proposed, as the Clique Percolation Method (CPM) in~\cite{palla2005}, and the algorithm in~\cite{Lanci2009}, based on a fitness function. Both of them find overlapping communities, and in the latter, a different notion of community as a {\em natural community} arose. The {\em natural community} of a vertex is a locally-computed set, and its size depends on a resolution parameter $\alpha$.

It has also been observed that the resolution limits for modularity found in~\cite{fortunato:rlicd} are particularly common in heterogeneous graphs with heavy-tailed community sizes and vertex degree distributions (see~\cite{Santo201075}, section VI.C). In these graphs, small communities will often be masked into larger ones by modularity maximization techniques when they are interconnected just by a few links. 

In order to detect the communities we define a {\em fitness function} following the ideas in~\cite{Lanci2009}. After analyzing the role of the resolution parameter $\alpha$ in these functions, we propose a {\em uniform fitness growth process} which scans the whole graph and whose parameter is updated dynamically. Then, we extract a community partition from the output of this process. The details of our method are described in sections~\ref{sec:method} and~\ref{sec:3stages}, and the algorithmic complexity is discussed in section~\ref{sec:complexity}.

In section~\ref{sec:test_networks} we use a benchmark developed in~\cite{Lanci2008} to build a dataset of heterogeneous networks. The results that we obtained show an important improvement using our fitness growth process when compared to the global modularity maximization techniques, which suggests that local methods may outperform global ones in these cases. In order to discuss this conjecture, we propose a correlation-based measure of community structure and use it to visualize the differences in performance between the two methods, giving a possible explanation.

As a measure for comparing community structures, \cite{danon_mi} proposed using the {\em normalized mutual information}. We shall use it in order to make comparisons with global methods and with community structures known {\em a priori}. We also apply the algorithm to real networks and show the results. Finally, we discuss the robustness (repeatability of the results) of our process.


\section{Our method}\label{sec:method}

\cite{Lanci2009} defines a process based on a fitness function with a resolution parameter $\alpha$ such that, given a set $C \subset V$:
\[
f(C)=\frac{k_{in}}{(k_{in}+k_{out})^\alpha}
\] 
where $k_{in}$ is the number of edges that join vertices in $C$,
and $k_{out}$ is the number of edges that join some vertex in $C$ to some vertex not in $C$. Applying this process to any vertex $v$, the {\em natural community of v} is obtained. In some way, the resolution parameter $\alpha$ is related to the natural community size.

Starting with a community made up by the seed vertex $v$,
their algorithm proceeds by stages,
where in each stage the steps are: 1) select a vertex whose addition increments the fitness function, and add it to  the actual community;
2) delete from the actual community all the vertex whose deletion increments
the fitness function.

The algorithm stops when, being in stage 1, it finds no vertex to add.
Step 2 is time-consuming, and usually very few vertices are deleted,
but it is necessary due to the local, vertex-by-vertex nature of the analysis.
The authors called the final result of the algorithm the natural community associated to $v$.

In order to obtain a covering by overlapping communities, they select a vertex at random,
obtain its natural community, select a vertex not yet covered at random, obtain its natural community, and so on until they cover the whole graph.

In all this process, the resolution parameter $\alpha$ of the fitness function is kept fixed. The authors perform an analysis in order to find the significant values of $\alpha$.

Our contribution extends that work to define a {\em uniform growth process}. This process covers the whole graph by making a course throughout its communities. We modify the {\em fitness function} $f(C)$ and analyze the role of $\alpha$ in the termination criteria for the process. Then we propose an algorithm for increasing the fitness function monotonically while traversing the graph, dynamically updating the parameter. Finally, a {\em cutting technique} divides the sequence of vertices obtained by the process, in order to get a partition into communities.

\subsection{Previous definitions}

We shall deal with simple undirected graphs $G=(V,E)$, with $n=|V|$ vertices and $m$ edges (here $|.|$ denotes the cardinal of a set).
To avoid unnecesary details, we assume that $E\subset V\times V$
is such that $(v,w)\in E$ implies that $(w,v)\in E$.

We set $\delta_E(v,w)=1$ if $(v,w)\in E$, $\delta_E(v,w)=0$ in the other case.
We have then the following expression for the degree of a vertex $v$
\[
\deg(v)=\sum_{w\in V} \delta_E(v,w)\enspace .
\]
Thus, $|E|=\sum_{w\in V}{\deg(w)}=2m$.
We shall use two measures, $m_V$ and $m_E$,
the first one on $V$ and the second one on $V\times V$.
Given $C\subset V$,
\[
m_V(C)=\sum_{v\in C}\deg(v)/|E|
\]
is  the normalized sum of the degrees of the vertices in $C$.
Given $D\subset V\times V$,
\[
m_E(D)=\sum_{(v,w)\in D}\delta_E(v,w)/|E|\enspace .
\]
Notice that when $C_1, C_2\subset V$ are mutually disjoint,
$m_E(C_1\times C_2)$ is the normalized cut between $C_1$ and $C_2$.
The $cut(C_1,C_2)$ is, in this case, the set of pairs $(v,w) \in E$ such that $v \in C_1$ and $w \in C_2$.
Notice also that $m_V$ is the marginal measure of $m_E$,
and that these measures are in fact probabilities. 
For $C\in V$, we shall denote for simplicity $m_E(C)=m_E(C\times \bar{C})$, where $\bar C=V \setminus C$.

Let $C\subset V$, and $v\in V$. We denote 
\[
ki_C(v)=\sum_{w\in C} \delta_E(v,w)
\]
and 
\[
ko_C(v)=\sum_{w\not\in C} \delta_E(v,w)\enspace .
\]
Thus $ki_C(v)$ is the number of vertices in $C$ joined to $v$,
and $ko_C(v)$ is the number of vertices not in  $C$ joined to $v$;
of course $ki_C(v)+ko_C(v)=\deg(v)$.

We shall also use $ski(C)=\sum_{v\in C} ki_C(v)$,
and $sko(C)=\sum_{v\in C} ko_C(v)$\enspace .

\subsection{A growth process}\label{sec:lanci_process}

Consider a fitness function $f$, associating to each $C\subset V$
a real number $f(C)$.

Given $v\in V$, we shall consider a growth process
for $f$ with seed $v$:
it consists of a double sequence 
\[
D_{00},D_{10},\ldots,D_{1k_1},\ldots,D_{a0},\ldots,D_{ak_a},\ldots,D_{b0},\ldots,D_{bk_b}
\] 
of subsets of $V$. Thus, for each $a$ such that $0\leq a\leq b$, we have a subsequence  $D_{a0},\ldots,D_{ak_a}$ ($a,b \in \mathbb{N}$).
\begin{itemize}
\item
$D_{00}=\{v\}$, $k_0=0$.
\item
For $a\geq 0$, $D_{(a+1)0}=D_{ak_a}$ and $D_{(a+1)1}$ is obtained from $D_{(a+1)0}$ by adding to it one vertex such that
$f(D_{(a+1)1})>f(D_{(a+1)0})$.
\item
For $k\geq 1$, $D_{a(k+1)}$ is obtained from $D_{ak}$ by elimination of a vertex
(different from the seed vertex $v$),
such that $f(D_{a(k+1)})>f(D_{ak})$.
\end{itemize}
In addition, we assume that for each $a>0$, there is no vertex $w\in D_{ak_a}$
such that its elimination induces an increase in $f$,
and that there is no vertex out of $D_{bk_b}$ whose addition induces
an increase in $f$.  
Alternatively, we may describe the process
by $v+s_1w_1+\ldots s_rw_r$,
where the \emph{signs} $s_i$ ($1$ or $-1$) determine whether the vertex $w_i$
is added or eliminated in this step, for example
$v+w_1+w_2+w_3+w_4-w_5+w_6$ means that in the first four steps
we added $w_1,w_2,w_3,w_4$, in the fifth step we eliminated $w_5$
(which of course must be equal to some of the previously added
vertices) and in the sixth step we added $w_6$.
\subsection{Concrete cases}
For $C\subset V$, consider
$m_V(C), m_E(C)$,
which we shall abbreviate $m_V, m_E$ when there is no place for ambiguity.
Recall that $m_V$ is the normalized sum of the degrees of the vertices in $C$,
and $m_E$ is the normalized cut defined by $C$.

We shall deal with two parametric families of fitness
functions, with a real parameter $t>0$:
\[
L_t=\frac{m_V-m_E}{m_V^{1/t}}
\]
and
\[
H_t=m_V(1-m_V/2t)-m_E\enspace .
\]

The first of these families is equivalent to
the one used by the authors in \cite{Lanci2009},
with $\alpha=1/t$.

\subsection{A differential analysis}
Let $C\subset V$, and $w\in V$.
Suppose that we are to add $w$ to $C$, if $w\not\in C$,
or to eliminate $w$ from $C$, if $w\in C$,
obtaining in either case a new set $C'=C\pm w$.
Let us denote $\Delta m_V=m_V(C')-m_V(C), \Delta m_E=m_E(C')-m_E(C)$, and $s, t>0$ two fixed values of the parameter.
Then we have the following approximate expression for the difference quotient of $L_t$,
\[
\frac{\Delta L_t}{\Delta m_V}\approx L_t'=\frac{1}{m_V^{1/t}}\left(1-\frac{\Delta m_E}{\Delta m_V}-\frac{L_1}{t}\right)\enspace .
\]

For the difference quotient of $H_t$ we obtain
\[
\frac{\Delta H_t}{\Delta m_V}\approx H_t'=\left(1-\frac{\Delta m_E}{\Delta m_V}-\frac{m_V}{t}\right)\enspace .
\]

Notice then the following relations
\begin{eqnarray}
H'_t&=&H'_s+\frac{t-s}{ts}m_V\label{derivadaH}\\
m_V^{1/t} L'_t&=&m_V^{1/s} L'_s+\frac{t-s}{ts}L_1\label{derivadaL}\\
H'_t&=&m_V^{1/t}L_t'+(L_1-m_V)/t\label{derivadaHL}
\end{eqnarray}

Equation \ref{derivadaH} shows us
that if $t>s$ and $H'_s>0$, then $H'_t>0$,
which means that if the vertex $w$ is a candidate for addition (elimination)
to $C$ (from $C$) for the $H_s$ process, it is also
a candidate for addition (elimination) for the $H_t$ process.

Equation \ref{derivadaL} shows us analogously
that if $t>s$ and $L'_s>0$, then $L'_t>0$,
which means that if the vertex $w$ is a candidate for addition (elimination)
to $C$ (from $C$) for the $L_s$ process, it is also
a candidate for addition (elimination) for the $L_t$ process.

This shows that the parameter $t$ does not play an essential role
during the growth process for $H_t$ or $L_t$, but merely establishes the 
\emph{termination criteria}.

Equation \ref{derivadaHL} shows
a delicate fact:
If a vertex  $w$ is a candidate for addition (elimination) for the  $L_t$ process,
and $m_V<L_1$ (this is usually true,
notice that when $m_V>L_1$, 
$m_E>m_V(1-m_V)$, which contradicts the notion of community,
because the second term would be the mean of the first one
if the vertices were to be selected randomly)
then it is a candidate for addition (elimination) for the $H_t$ process.
Thus, both processes are essentially equivalent,
their difference lying in the termination criteria.
In exceptional cases, communities obtained with
the $H_t$ fitness functions are bigger than those 
obtained with the $L_t$ fitness functions.

Of course, there are approximations involved,
so that our previous comments are rough and qualitative:
our experience testing both fitness functions
confirms them. 

\subsection{Natural communities}

The following is a formalization of the procedure described in~\cite{Lanci2009} to obtain the natural community of a vertex $v$, generalized for any fitness function.

\begin{algorithm}[H]
  \footnotesize
  \caption{Natural communities}\label{natural_comm_algorithm}
  \SetLine
  \KwIn{A graph $G=(V,E)$, a fitness function {\em f}, a vertex $v \in V$}
  \KwOut{A growth process $D_{00},D_{10},\ldots,D_{a0},\ldots,D_{ak_a},\ldots,D_{b0},\ldots,D_{bk_b}$}
  \Begin{
     $D_{00} = \{v\}$ \\
     $m = 0$ \\
     \While {there exists $w$ out of $D_{m0}$ such that $f(D_{m0}+w)>f(D_{m0})$}{
	$D_{m1}=D_{m0}+w$ \\
	$k=1$ \\
	\While {there exists $w\in D_{mk}, w\not = v:f(D_{mk}-w)>f(D_{mk})$}{
	  $D_{m(k+1)}=D_{mk}-w$ \\
	  $k=k+1;$

        }
	$D_{(m+1)0}=D_{mk}$ \\
	$m=m+1$
      } \label{endloop1}
  }
\end{algorithm}

The output of this ``algorithm'' is a growth process for $f$,
$v+w_1+w_2\pm w_3\pm\ldots\pm w_{r-1}+w_r$,
such that there is no  $w$ not in $D_{r0}$ with $f(D_{r0}+w)>f(D_{r0})$.
Each $D_{j0}, 0\leq j \leq k$ satisfies that there is no
$w\in D_{j0}, w\not =v$, such that $f(D_{j0}-w)>f(D_{j0})$.
$D_{r0}$ is a possible ``natural community'' with seed $v$.

{\em Remark:}
Notice that the preceding prescription is not complete,
because both the $w$ that we choose to add,
as well as the $w$ that we choose to eliminate,
depend upon a criterion that we do no fix.

\subsection{Uniform growth processes}
In the previous Section we have described a method to obtain a natural
community with seed $v$ and fitness function $f$. Applying this with $f=H_t$ 
and fixed $t$, for different values of $t$ we obtain different communities.
Although it is not strictly true that ``the bigger the $t$, the bigger the community'',
we have noticed in our differential analysis that this is essentially the case.
Thus, it is reasonable to wonder whether it is possible to obtain all these communities with a unique process, starting with the smallest ones and proceeding with the biggest ones. The answer is affirmative, as we shall see now. 
 
Let us assume that we have
our parametric family of fitness functions $H_t:0<t$.
Given $C$ and $w\in V$ such that $ki_C(v)>0$, there 
always exists $t_c=t_c(C,w)>0$ such that
$H_{t_c}(C+w)=H_{t_c}(C)$.
Indeed, we have:
\begin{eqnarray*}
H_{t}(C+w)&=&(m_V+\Delta m_V)(1-(m_V+\Delta m_V)/2t)-(m_E+\Delta m_E)\\
&=&
m_V(1-m_V/2t)-m_E-\frac{\Delta m_V}{t}(m_V+\Delta m_V/2)+\Delta m_V-\Delta m_E\\
&=&
H_{t}(C)-\frac{\Delta m_V}{t}(m_V+\Delta m_V/2)+\Delta m_V-\Delta m_E
\end{eqnarray*}
and it follows that
\[
t_c=\frac{\Delta m_V(m_V+\Delta m_V/2)}{\Delta m_V-\Delta m_E}
\]
satisfies our exigencies.
We also see that
\[
\Delta H_t=-\frac{\Delta m_V}{t}(m_V+\Delta m_V/2)+\Delta m_V-\Delta m_E
\]
and it follows that $\Delta H_t>0$ when $t>t_c$ and $w\not\in C$,
and that $\Delta H_t>0$ when $t<t_c$ and $w\in C$.

Let $v+\sum_{i=1}^M s_iw_i$
be an algebraic expression with the previously introduced meaning,
where of course we assume that each time that we eliminate a vertex,
that vertex had previously been added.
Let $C_0=v$ and for $r>0$, $C_r=v+\sum_{i=1}^r s_iw_i$. We assume that for each $r$, $0\leq r<M$, 
$ki_{C_r}(w_{r+1})>0$.
We shall consider values $0=t_0,t_1,\ldots, t_r$ associated to this expression,
$t_r=max\{t_{r-1},t_c(C_{r-1},w_r)\}$ when $s_r=1$,
$t_r=t_{r-1}<t_c(C_{r-1},w_r)$ when $s_r=-1$.
Thus, $t_0,\ldots,t_r$ is a non-decreasing sequence,
and  $C_0,\ldots,C_r$  is a growth process for $H_t$
if $t>t_r$. We call $C_0,\ldots,C_M$ a \emph{uniform growth process} for $H$.

\begin{algorithm}[H]
  \footnotesize
  \caption{A growth process for $H$}\label{ht_algorithm}
  \SetLine
  \KwIn{A graph $G=(V,E)$, a vertex $v \in V$}
  \KwOut{A growth process for $H$: $D_{00},D_{10},\ldots,D_{a0},\ldots,D_{ak_a},\ldots,D_{b0},\ldots,D_{bk_b}$}
  \Begin{
     $D_{00} = \{v\}$ \\
     $t_a = 0$ \\
     $m = 0$ \\
     \While {there exists $w$ not in $D_{m0}$}{
let $w_0$ be such that $t_c(D_{m0},w_0)=\min_{w\not\in D_{m0}}(t_c(D_{m0},w))$\\
$t_a=\max\{t_a,t_c(D_{m0},w_0)\}$\\
	$D_{m1}=D_{m0}+w_0$ \\
	$k=1$ \\
	\While {there exists $w\in D_{mk}, w\not = v:t_c(D_{mk},w)>t_a$}{
	  $D_{m(k+1)}=D_{mk}-w$ \\
	  $k=k+1;$
        }
	$D_{(m+1)0}=D_{mk}$ \\
	$m=m+1$
      } \label{endloop2}
  }
\end{algorithm}

The output of this ``algorithm'' is a uniform growth process for $H$, which ends by covering the whole graph. The successive truncations of the sequence thus obtained are natural communities for $v$ at different resolutions. In the sequel we assume -with empirical evidence- that these natural communities are made up of small subcommunities, which are inserted one after another during the growth process. The following section explains how to detect these communities.

\section{Extracting the communities in three stages}\label{sec:3stages}

The previous section described the growth process, which outputs a sequence $C_r=v+\sum_{i=1}^r s_iw_i$. Some vertices of the graph may be inserted, removed and later reinserted during this process. So as a first step we filter the sequence to generate a new one which only keeps the last insertion of each vertex. In this way we obtain a subsequence $\mathcal{S}$ of the original one, such that each vertex appears once and only once throughout it. Now, as the growth process tends to choose the vertices by their strong linkage to the natural community built so far, we state that two consecutive vertices in the sequence either belong to the same
community or either are border vertices. Considering that the first case is the
most frequent, an algorithm is needed in order to cut that sequence into
communities. This section presents our approach in three stages to obtain
the final partition of the graph. Briefly, the first stage turns the sequence of
vertices into a sequence of communities. It makes use of a
 division criterion defined by a function $R(v)$ in order to decide if a vertex $v$ will stay in the
same community as the previous vertex in the sequence or it will start a new community.
The second stage will join consecutive communities in order to improve the community structure, and the 
last stage will move individual vertices from one community to another.

\subsection{Stage One: Making cuts in the process}

In this first stage we divide the sequence $\mathcal{S}$ to obtain a list of communities $\mathcal{C}=(C_1, C_2, ..., C_{M})$. These communities are composed by vertices which are consecutive in the sequence. The cuts are made by observing the behavior of the function

\begin{equation}
R(w)=\frac{ki_{S(w)}(w)-ko_{S(w)}(w)}{ki_{S(w)}(w)+ko_{S(w)}(w)} \enspace ,
\end{equation}
where $S(w)$ are the sublists of $\mathcal{S}$, from the first vertex in the sequence, up to $w$.

Figure 2 sheds some light on why this function is useful to identify ``subcommunities'', i.e., elementary groups which will later take part in the final communities.

\begin{figure}
\begin{center}
\includegraphics[width=16cm]{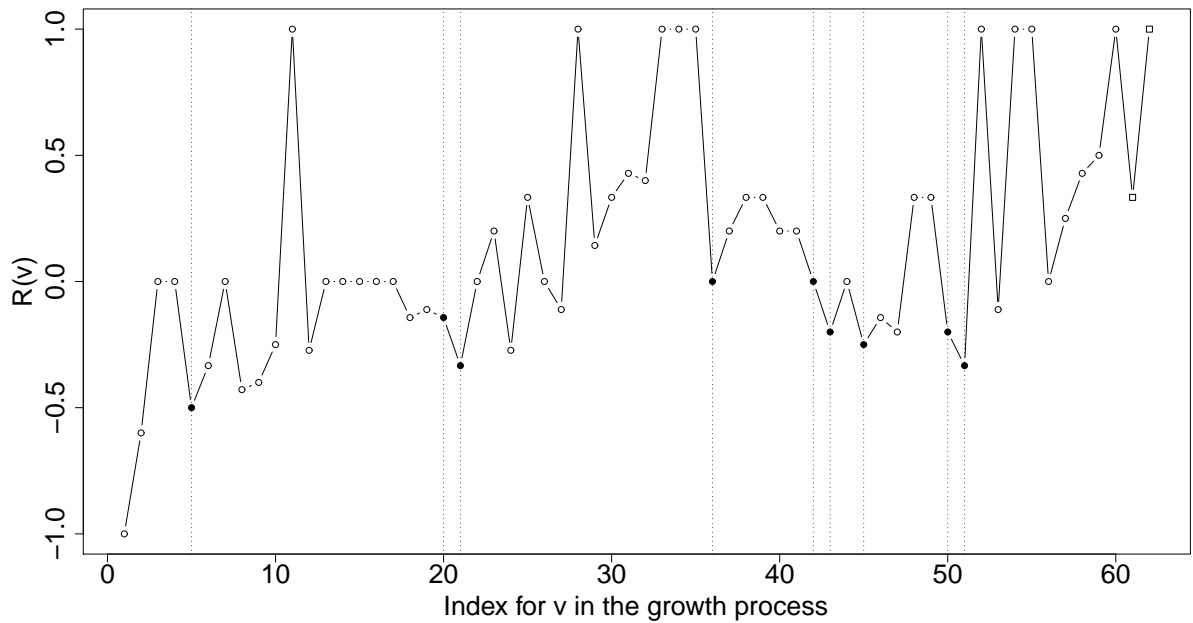}
\caption{The cuts in the growth process for the dolphins social network~\cite{Lusseau_Newman_2004}. The cut vertices (in black) are: 44, 36, 3, 0, 39, 7, 1, 41, 57. 
\label{dolphins_cuts}}
\end{center}
\end{figure}

In fact, what happens is that when the process leaves a subcommunity of strongly connected vertices and adds any vertex from outside, there is a decay in the function value, due to the relatively scarce number of connections between the subcommunity and the new vertex. Figure~\ref{dolphins_cuts}, obtained processing the {\em dolphins} network~\cite{Lusseau_Newman_2004}, shows a clear decay in position $36$ when the process jumps between the two known communities~\cite{newman:faecsin}.

\begin{figure}
\begin{center}
\includegraphics[width=6.8cm,angle=270]{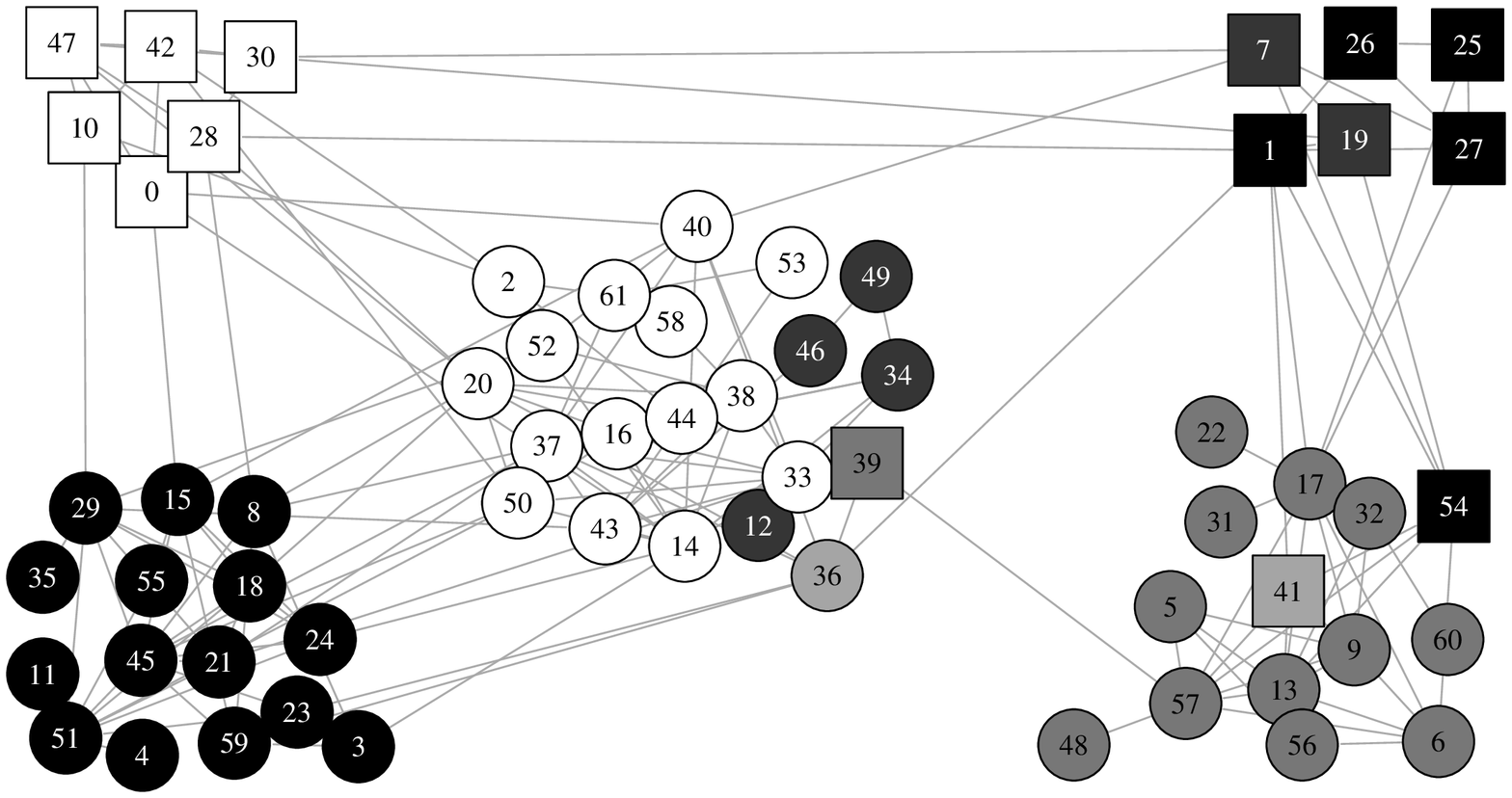}
\includegraphics[width=6.8cm,angle=270]{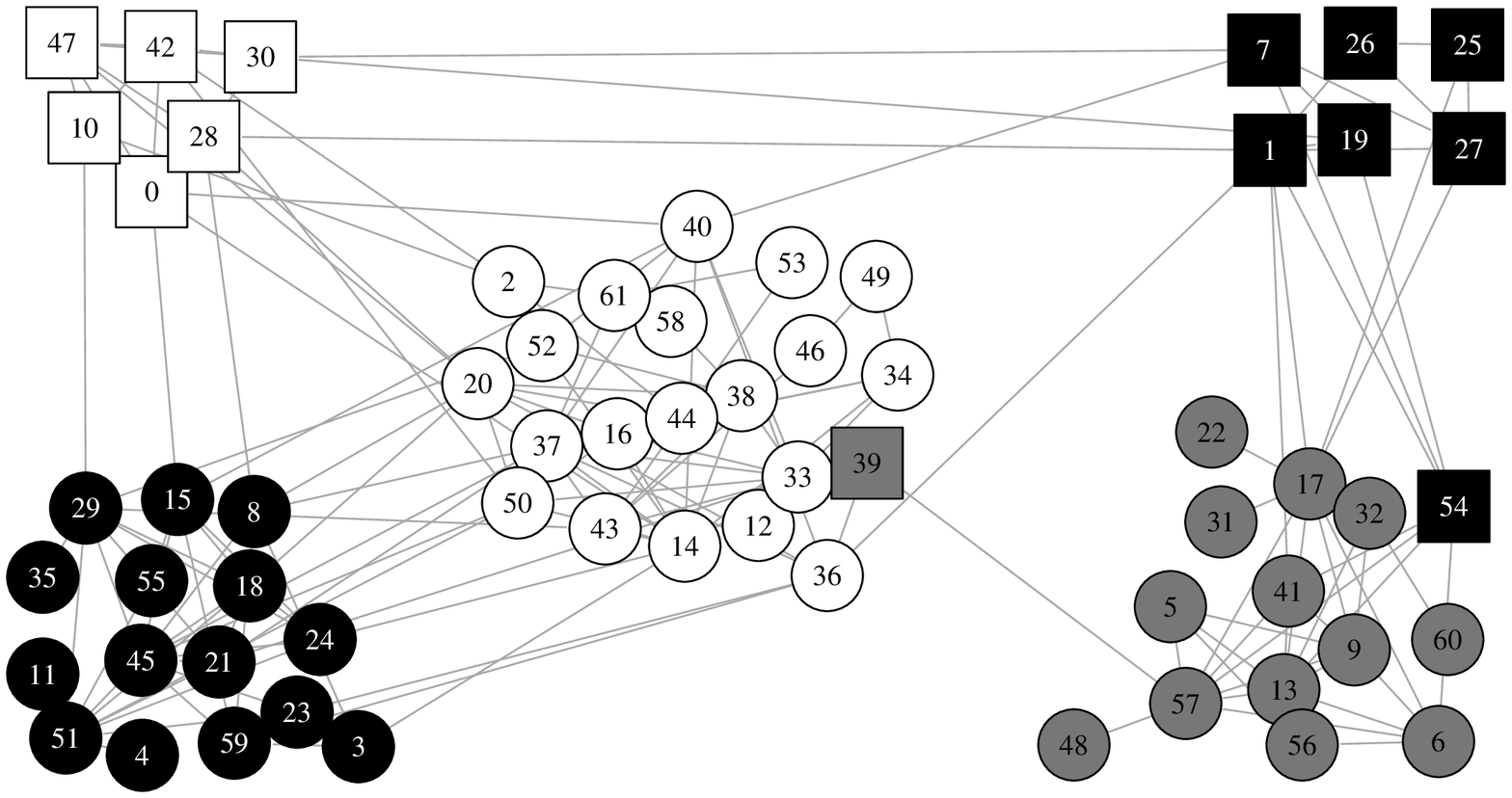}
\includegraphics[width=6.8cm,angle=270]{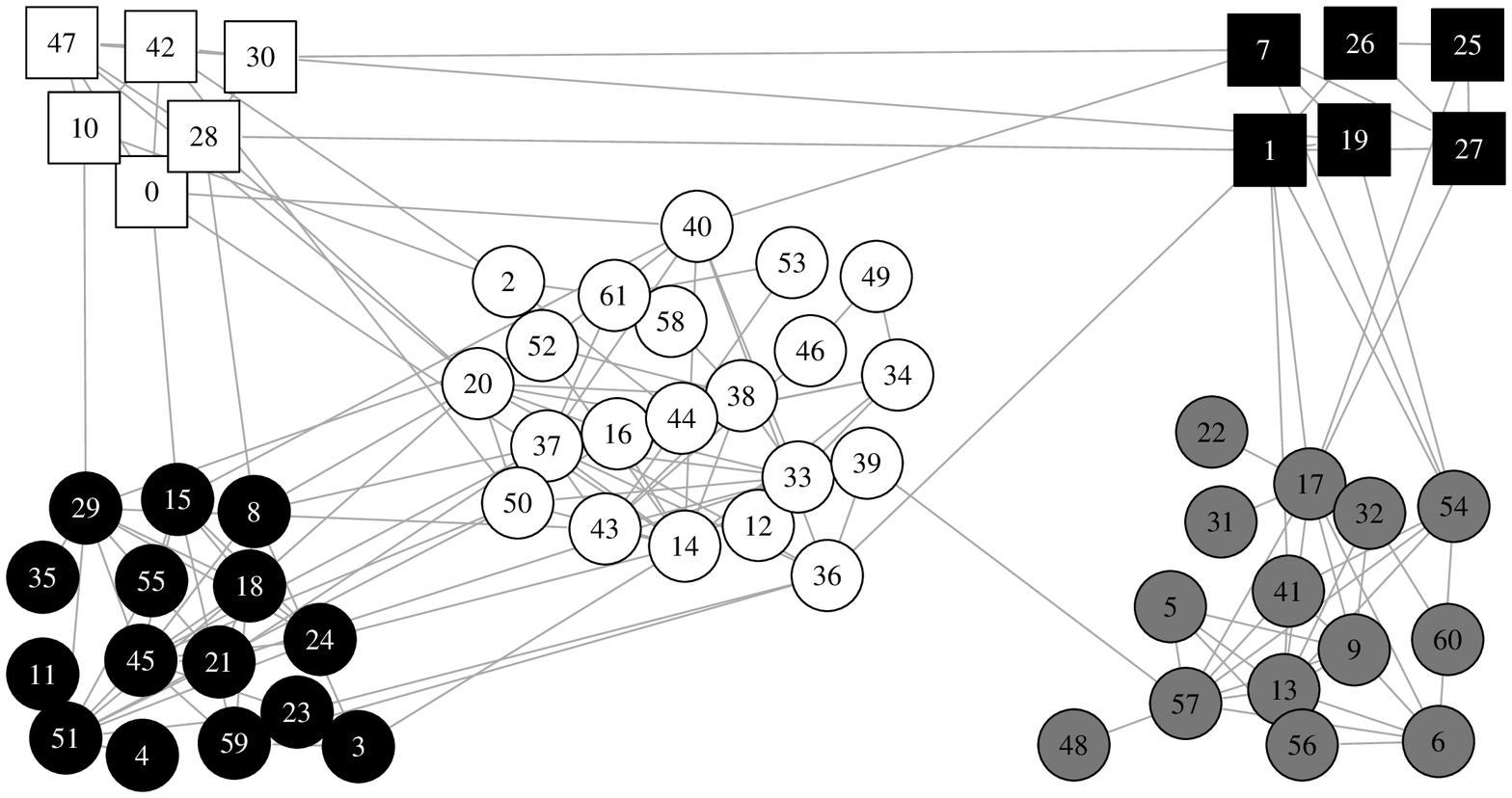}
\caption{The three stages of the algorithm in the {\em dolphins}
network. The vertices were positioned according to their communities after
the third stage. Picture generated with the {\em igraph} package for
{\em R}~\cite{rpkg:igraph}. The picture for the first stage matches with the cuts in
Figure~\ref{dolphins_cuts} (from left to right) in the following way (initial vertex,
color and shape): 12, dark gray circles; 44, white circles; 36, light gray circles; 3, black circles; 0, white
rectangles; 39, gray rectangles, 7, dark gray rectangles; 1, black
rectangles; 41, light gray rectangles; 57, gray circles.  
\label{dolphins_s}}
\end{center}
\end{figure}

The $R(v)$ function cuts the sequence whenever it finds a minimum value which is smaller than the last minimum. This fact indicates that we have reached a valley between two bellies of the curve, which belongs to an inter-community area. This is quite an aggressive criteria, as sometimes frontier vertices may produce unnecessary cuts. This does not represent a problem, because this small communities taken from the border will be joined to their actual communities during the next stages. This is the case of the vertices in positions $36$, $39$ and $54$ in Figure~\ref{dolphins_s}. This figure illustrates the three stages for the dolphins network.

\subsection{Stage Two: Joining successive sets to get communities}

In this step we join consecutive subcommunities $(C_i, C_{i+1})$ from stage 1, based on the following criteria: when $cut(C_i,C_{i+1})>ski(C_i)$ or $cut(C_i,C_{i+1})>ski(C_{i+1})$ (which means that the subcommunity has more connections to the other one than to itself), then the subcommunities are merged and form a new community $C'_i$. The step finishes when no more consecutive subcommunities can be joined.

\subsection{Stage Three: Reclassifying vertices}

In order to correct the possible errors of the fitness growth process, we apply this last step, which is similar to the previous one, but with a vertex granularity: if any vertex $w$ has more connections to some other community $C_j$ than to the one it belongs to, then the vertex is moved to $C_j$. When this stage finishes every vertex is more attached to its own community than to any other, which is quite a strong condition on community membership.

We sweep over all the vertices looking for misclassified ones, and when no vertex can be moved the algorithm stops. We have observed a fast convergence and stabilization of this stage in all the test networks that we used. During the first run, all vertices tend to move to their right community, and in the second and third runs the amount of moving vertices sharply decreases.

\section{Algorithmic Complexity}\label{sec:complexity}

In this section we provide complexity bounds for the growth process and for the three stages. We shall use the notation $N(v)$ for the neighborhood of $v$ (the set of vertices which have an edge with $v$). Similarly, $N(C)$ will denote the set of communities whose vertices have at least one neighbor in $C$. Finally, we call $d_{\max}=\max\{\deg(v), v\in V\}$.

{\em Growth process.} The growth process is a sequence of vertex insertions interleaved with some eliminations. During all our experiments, we verified that the eliminations are scarce and they do not affect the order of complexity of the process. So we shall analyze the complexity for a growing process with no eliminations, such that the community size grows linearly from 1 to $n$ on each step. Let's consider step $k$: we must analyze the inclusion of all the community neighbors, that is, all the vertices outside $C$ which have some neighbor in $C$; as $k$ vertices are inside $C$, the outsiders can be bounded by $n - k$. For each of them we evaluate $t_c(C, w)$. This implies computing $\Delta m_V$ and $\Delta m_E$: $\Delta m_V$ comes from the vertex degree, while $\Delta m_E$ is related with $ki_C$ and $ko_C$. So this computation is direct and does not depend on the size of the network. The minimum $t_c(C, w_i)$ wins and $w_i$ in inserted into the community $C$. The last step consists on updating the $ki$ and $ko$ for the neighbors of $w$, and for $w$ itself. For each of them we shall increase $ki$ by $1$ and decrease $ko$ by the same amount. The complexity of this last step is then $|N(w)|+1$.

Expanding the analysis for step $k$ to all the process, we get: $\sum_{k=1}^{n}{(n-k)+|N(w)|+1} \leq n^2 + n \cdot d_{max} + n$. This makes a complexity of $O(n^2)$.

{\em Stage 1.} In the {\em cutting algorithm} the process is run through only once, from the begin up to the end, and for vertex $v_i$, the cut decision is made based on $R(v_{i-1})$, $R(v_i)$ and $R(v_{i+1})$, where $i$ refers to the position of the vertex in the growth process. The complexity here is $O(n)$.

{\em Stage 2.} For the merge of communities which are consecutive in the process, we need a matrix with all the cuts $cut(C_i, C_j)$, and also the values of $ski$ and $sko$ for each community. In order to precompute all this, we must consider each edge in the network, so it has a cost of $O(m)$, and requires a memory of $O(|C|^2)$ (in order to build the adjacency matrix of communities). Now, after building this structure, we start merging consecutive communities. We can bound the number of merges with $|C|$, and for each merge we analyze all the possibilities, i.e., all the pairs $(C_i,C_{i+1})$, which totalize $(|C|-1)$. Evaluating the convenience of joining $C_i$ and $C_{i+1}$ is $O(1)$, as it only involves the pre-computed values of $ski$ and $sko$. So the selection of the best merge is $O(|C|)$. Finally, the update of the cuts $cut(C_i, C_j)$ for the neighbor communities of both implies $|N(C_i)|$ accesses to the matrix. Updating the values of $ski$ and $sko$ is immediate. In conclusion, the merge complexity is $O(|C|)$ and the number of merges is bounded by $|C|$. As $|C|$ is bounded by $n$, the cost of stage 2 is $O(n^2)$.

{\em Stage 3.} Here we analyze each pair $(v, C)$, where C is a community such that its vertices have one or more links to $v$. In order to decide if we move $v$ to $C$, we use an ordered record of the cuts $cut(v, C)$. Building the record at the beginning costs $O(m)$, just as in Stage 2. Then, we analyze all vertices ($O(n)$) to find the best community for each of them, and if we move the vertex, we must update the record, with a cost of $\deg(v)$. Now, this makes a complexity of $O(m + A \cdot n \cdot \deg(v))$, where $A$ is the number of traverses over all the vertices. Bounding this number with a fixed value -based on empirical observations-, the complexity is also $O(n^2)$.

\section{Results and Data Analysis}\label{sec:test_networks}

In this section we exhibit the results of our local method applying it to {\em (i)} a benchmark of heterogeneous networks, {\em (ii)} real networks of different sizes, {\em (iii)} random networks. We develop a brief explanation about mutual information as a metric in~\ref{sec:mutual_inf}, and in~\ref{sec:corr_based} we propose a correlation-based measure which shall be useful to understand the limits of global methods. Finally we show that the algorithm is robust for large networks with a well-defined community structure.

\subsection{Mutual Information}\label{sec:mutual_inf}

For the purpose of comparing different community structures, we used the {\em normalized mutual information}~\cite{danon_mi}. In order to define it in terms of random variables, we consider the following process: we pick a vertex $v$ at random from $V$ with a uniform distribution, and define the variable $X$ related with partition $\mathcal{C}_1$. This variable assigns to each vertex the subindex of the community it belongs to.
Clearly, the distribution of $X$ is

\begin{equation}
\prob[X=i]=p_{i} = \frac{|C_i|}{|V|}, \enspace ,
\end{equation}
where $i = 1, 2, ..., |\mathcal{C}_1|$.
The entropy of $\mathcal{C}_1$ can now be defined as:

\begin{equation}
H(\mathcal{C}_1) = -\sum_{i=1}^{|\mathcal{C}_1|}{p_i \cdot log \left(p_i\right)}\enspace .
\end{equation}

If we introduce a second partition $\mathcal{C}_2$ with its related variable $Y$ under the same process, then the joint distribution for $X,Y$ is

\begin{equation}
\prob[X=i,Y=j]=p_{ij} = \frac{|C_i \cap C_j|}{|V|}, \enspace ,
\end{equation}
where $i = 1, 2, ..., |\mathcal{C}_1|$, $j = 1, 2, ..., |\mathcal{C}_2|$.
In these terms, the normalized mutual information is expressed as:

\begin{equation}
NMI(\mathcal{C}_1,\mathcal{C}_2) = -2 \cdot 
\frac
{\sum_{i=1}^{|\mathcal{C}_1|}{\sum_{j=1}^{|\mathcal{C}_2|}{p_{ij} \cdot log \left(\frac{p_{ij}}{p_i \cdot p_j}\right) }}}
{\sum_{i=1}^{|\mathcal{C}_1|}{p_i \cdot log\left(p_i\right)}+\sum_{j=1}^{|\mathcal{C}_2|}{p_j \cdot log\left(p_j\right)}}\enspace ,
\end{equation}
where ${\sum_{i=1}^{|\mathcal{C}_1|}{\sum_{j=1}^{|\mathcal{C}_2|}{p_{ij} \cdot log \left(\frac{p_{ij}}{p_i \cdot p_j}\right) }}}=MI(\mathcal{C}_1,\mathcal{C}_2)$ is the mutual information. The following equality holds:

\begin{equation}
MI(\mathcal{C}_1,\mathcal{C}_2) = H(\mathcal{C}_1) + H(\mathcal{C}_2) - H(\mathcal{C}_1, \mathcal{C}_2) \enspace ,
\end{equation}
where $H(\mathcal{C}_1, \mathcal{C}_2)$ is the joint entropy.
$NMI(\mathcal{C}_1,\mathcal{C}_2)$ falls between $0$ and $1$, and gives an idea of the similarity between partitions in terms of the information theory, i.e., in terms of the information about $\mathcal{C}_1$ that lies in $\mathcal{C}_2$, or vice versa.

The inherent idea is that a partition $\mathcal{C}$ of a graph gives us some information relative to the classification of vertices into groups. This amount of information is measured by its entropy, $H(\mathcal{C})$.

In fact, the denominator in $NMI(\mathcal{C}_1,\mathcal{C}_2)$ together with the $-2$ constant represent a normalization by the average entropy of the partitions, $\frac{H(\mathcal{C}_1)+H(\mathcal{C}_2)}{2}$. A normalized mutual information of $1$ implies that the partitions are coincident.

\subsubsection{Normalizations and triangular inequalities}\label{sec:mutual_inf_transitivity}

We remark that other normalizations of the mutual information also exist, like:

\begin{equation}
NMI_2(\mathcal{C}_1,\mathcal{C}_2) = \frac{MI(\mathcal{C}_1, \mathcal{C}_2)}{H(\mathcal{C}_1, \mathcal{C}_2)}
\end{equation}
which has the advantage that $1-NMI_2$ is a metric~\cite{vinh}. Although we consider it more correct to use this normalization, we shall hold to the first one for the purpose of comparison with other works in the literature. Anyway, we were able to find a transitivity property on $NMI$ too (we shall call it $NMI_1$ here). In fact, observing that:

\begin{equation}
\frac{2}{1-NMI_1(\mathcal{C}_1,\mathcal{C}_2)} = \frac{H(\mathcal{C}_1, \mathcal{C}_2)}{H(\mathcal{C}_1) + H(\mathcal{C}_2) - H(\mathcal{C}_1, \mathcal{C}_2)}
\end{equation}
\begin{equation}
\frac{1}{1-NMI_2(\mathcal{C}_1,\mathcal{C}_2)} = \frac{H(\mathcal{C}_1) + H(\mathcal{C}_2)}{H(\mathcal{C}_1) + H(\mathcal{C}_2) - H(\mathcal{C}_1, \mathcal{C}_2)}
\end{equation}

we can deduce a functional relationship between these two:

\begin{equation}
\frac{2}{1-NMI_1(\mathcal{C}_1,\mathcal{C}_2)} - \frac{1}{1-NMI_2(\mathcal{C}_1,\mathcal{C}_2)} = 1
\end{equation}

\begin{figure}[h]
\begin{center}
\includegraphics[width=7cm]{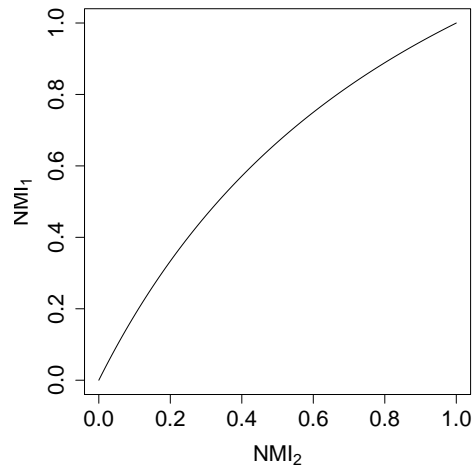}
\caption{\label{fig:hyperbola}Functional relationship between two normalizations of the mutual information: $NMI_1$ and $NMI_2$.}
\end{center}
\end{figure}

This relationship produces an hyperbole as in Figure~\ref{fig:hyperbola}. The good behavior of the function around $(1,1)$ assures that values of $NMI_1$ close to $1$ imply values of $NMI_2$ close to $1$ too. The transitivity of the metric implies that if $NMI_2(x,y) \geq 1 - \epsilon$ and $NMI_2(x,z) \geq 1 - \epsilon$, then $NMI_2(y,z) \geq 1 - 2 \epsilon$. Then, by the functional relationship, $NMI_1(y,z)$ will be somehow close to $1$ too.

In other words, if $NMI(\mathcal{C}_R,\mathcal{C}_1)$ is high and $NMI(\mathcal{C}_R,\mathcal{C}_2)$ is high, then $NMI(\mathcal{C}_1,\mathcal{C}_2)$ is also high. This result will be used in section~\ref{sec:robustness}, where $\mathcal{C}_R$ is a reference partition used to analyze our algorithm's robustness.

\subsection{Benchmarking with a set of heterogeneous networks}

\subsubsection{Benchmark description}\label{sec:bench_desc}

We evaluated our algorithm with a benchmark proposed in~\cite{Lanci2008}. We used their software to create sets of $10,000$ heterogeneous random graphs, with different power laws for the vertex degree distribution (exponent $\alpha$) and the community size distribution (exponent $\beta$), as well as different mixing parameters $\mu$.

We constructed graphs of 1,024 vertices, with $\langle \deg(v) \rangle = 10$ and $d_{\max} = 100$. Each set keeps a fixed value of $\alpha$ and $\beta$, while the mixing parameter $\mu$ moves between $0.05$ and $0.50$. Thus, it has $1,000$ graphs for each $\mu$, making a total of $10,000$ graphs.

We built $3$ sets, considering representative values of $\alpha$ and $\beta$ in heterogeneous networks.
\begin{itemize}
\item {\tt BENCH1}: $\alpha=1.2$, $\beta=3.0$
\item {\tt BENCH2}: $\alpha=1.8$, $\beta=1.2$
\item {\tt BENCH3}: $\alpha=2.0$, $\beta=2.0$
\end{itemize}

We also tested other pairings of $\alpha \in [1,3]$ and $\beta \in [1,3]$. {\tt BENCH1} turned out to be the best-case, {\tt BENCH2} the worst-case, and {\tt BENCH3} a mean-case. 

We have used this benchmark for different reasons: {\em (a)} it simulates real networks by generating heterogeneous distributions. These distributions provide greater challenges to the community discovery algorithms with respect to fixed-degree networks like the ones generated by the GN benchmark~\cite{girvan_2002}. For example, heterogeneous networks are subject to resolution limit problems when global methods are applied; {\em (b)} the parameters adjust tightly to the proposed values, the $\mu$ distribution following a roughly bell-shaped curve around the desired $\mu$; and {\em (c)} it has a low complexity, which makes it suitable to generate a big set of graphs.

\subsubsection{Obtained results}

\begin{figure}[h]
\begin{center}
\includegraphics[width=11cm,angle=270]{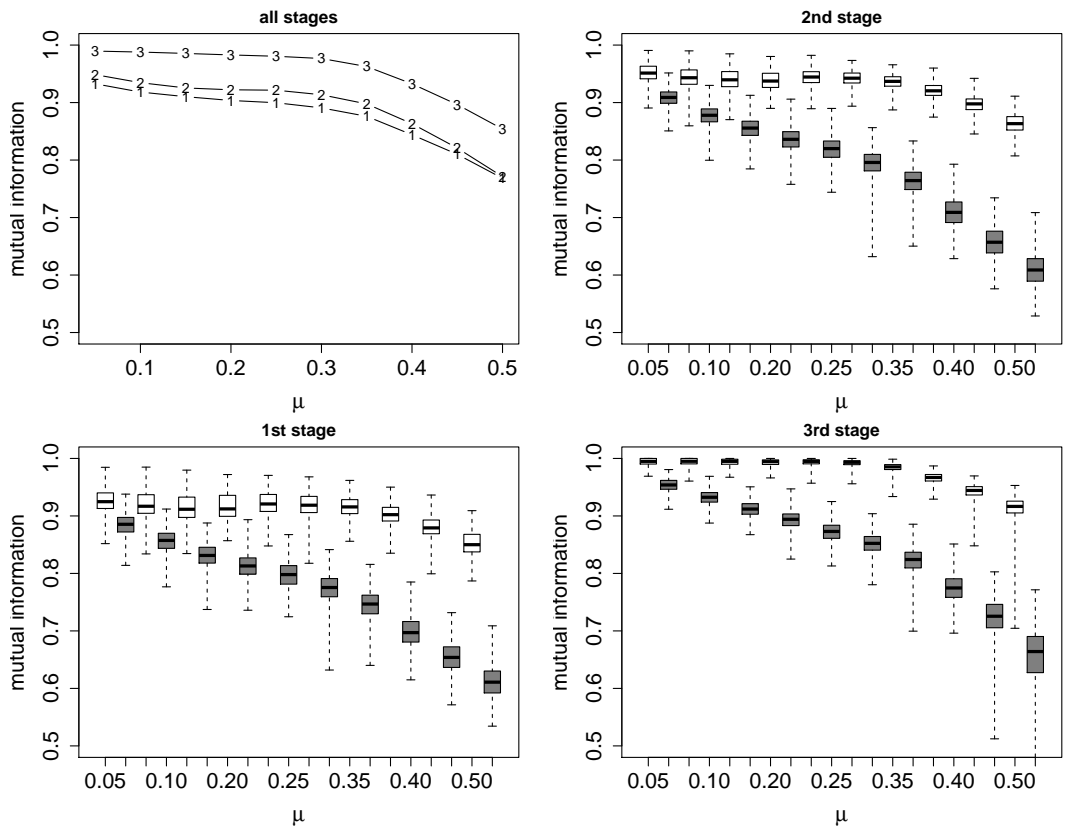}
\caption{\label{fig:mut-inf}Statistical analysis of the normalized mutual information between our partition and the communities known {\em a priori}, after each of the three stages of the community detection algorithm. These are results for {\tt BENCH1}, {\tt BENCH2} and {\tt BENCH3}, each of them consisting on 1,000 networks for each value of $\mu$, whose values range from $0.05$ to $0.50$. The plot in the upper-left corner is for {\tt BENCH3}, and represents median values of mutual information after each of the three stages. Each of the other plots compares {\tt BENCH1} {\em(white)} and {\tt BENCH2} {\em(gray)} for a different stage. $\mu$ varies from $0.05$ to $0.5$ in steps of $0.05$, but the boxplots are interlaced over the $x$-axis just for the sake of clarity.}
\end{center}
\end{figure}

As explained in section~\ref{sec:3stages}, the uniform growth process returns an ordered list of vertices, such that either two consecutive vertices are neighbors in
the same community, or else each of them belongs to its community border. Only after computing the first stage we get a partition that we can compare with the original one.
Figure~\ref{fig:mut-inf} analyzes the results of the three stages as a function of $\mu$, which is the most decisive parameter during the communities detection.
It displays the mutual
information between our partition
and the one issued from the benchmark, after the end of each stage. We
used the {\tt boxplot} command of the {\em R} statistical
software~\cite{Rmanual}. This command computes the quartiles for each $\mu$, displaying:
the median (second quartile); boxes representing the $3^{rd}$ and the $1^{st}$ quartiles; and 
whiskers which are placed at the extremes of data.
The plot in the upper left corner analyzes {\tt BENCH3}, and shows only
the medians for the three stages at the same time, for comparison purposes. The other plots are boxplots comparing {\tt BENCH1} and {\tt BENCH2}.

We observe that the results after the first stage on {\tt BENCH1} and {\tt BENCH3} are successful for a wide range of values of $\mu$, where the mutual information is larger than $0.9$. {\tt BENCH2} represents the worst-case, and greater values of $\mu$ make the mutual information decrease substantially. This is a typical behavior, and one of the reasons is that the first stage cuts the ordered
list in sets every time that it reaches a community border; as the borders are very fuzzy for big values of $\mu$, sometimes communities are
split in two or more. Then, it is the second stage the one which corrects
this problem, improving the last result in about $3\%$, being more
effective for lower values of $\mu$. 
Finally, the third stage makes a considerable gain in general, even for large
values of $\mu$. In fact, the mutual information improves more than
$10\%$ in the interval $\mu=[0.3,0.5]$. In the case of {\tt BENCH2} and $\mu=0.5$ the third stage improves the median but extends the range of values of the mutual information, reaching a minimum value of $0.2$.

\subsubsection{A comparison with a modularity-based method}

\begin{figure}[h]
\begin{center}
\includegraphics[width=16cm]{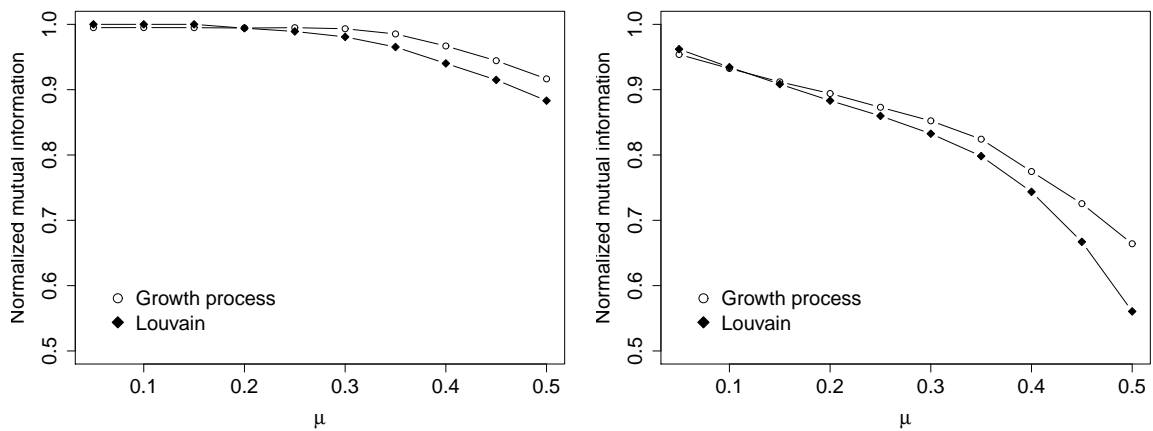}
\caption{\label{fig:comp-bgll} Comparison between our growth process and Louvain's modularity-based method. We consider the communities generated {\em a priori} by Lancichinetti's benchmark, and we use them as a reference partition for the comparison. The picture compares the mutual information for our growth process and for Louvain's method. The points represent median values for the 1,000 networks generated for each different $\mu$. (a) On the left, results for {\tt BENCH1}: $\alpha=1.2$, $\beta=3.0$. (b) On the right, results for {\tt BENCH2:} $\alpha=1.8$, $\beta=1.2$.}
\end{center}
\end{figure}

Figure~\ref{fig:comp-bgll} compares the partitions found with our growth process based on the $H$ fitness function, and a modularity based algorithm. We chose the Louvain algorithm~\cite{blondel:fuociln}, which is one of the most efficient modularity-based methods. The points represent median values for the 1,000 different networks in benchmarks {\tt BENCH1} and {\tt BENCH2}, varying the mixing parameter $\mu$. The reference partition is the one computed {\em a priori} by Lancichinetti's benchmark, from which the networks are generated. So when we mention the mutual information for the growth process we mean {\em the mutual information against the pre-computed communities}. The same holds for the mutual information for the Louvain algorithm.

We observe that our growth process represents a general improvement for the detection of communities in the benchmarks, and that the difference in performance increases for higher values of the mixing parameter $\mu$. This behavior will be argued in the next subsection.

\subsection{A correlation-based measure}\label{sec:corr_based}

Let $C_i, 1\leq i \leq k$ be a partition of $V$.
Consider the following random variables:
select a pair $(v,w)$ from $E$ at random and define
$L_i$ as a Bernoulli variable such that $L_i=1$ if $v\in C_i$. In the same way, we define $R_i$
as a Bernoulli variable such that $R_i=1$ if $w \in C_i$.
Thus, it follows that $\prob(L_i=1)=\prob(R_i=1)=m_V(C_i)$.
If $C_i$ is a community, we expect that $\prob(R_i=1|L_i=1)>\prob(R_i=1)$,
thus a sensible measure of the community quality is the correlation
$\rho_{ii}$, where
\[
\rho_{ij}=\rho(L_i, R_j)=\frac{m_E(C_i\times C_j)-m_V(C_i)m_V(C_j)}{\sqrt{m_V(C_i)m_V(C_j)(1-m_V(C_i))(1-m_V(C_j))}}
\]

Notice also that $\rho_{ij}>0$ means that joining $C_i$ to $C_j$
will give an increment in the usual Newman modularity $Q$,
and that $\rho_{ii}>0$ means that 
\[
\prob (R_i=1|L_i=1)>\prob (R_i=1)
\]
as expected. In~\cite{hamelin2010} the authors have studied the relationship between these coefficients $\rho_{ij}$ and modularity maximization, and when $\rho_{ij}>0$ they say that $C_i$ and $C_j$ are mutually {\em submodular}. This simply means that this pair of communities would be usually joined by agglomerative modularity maximization techniques, because their union increases modularity.

\begin{figure}[h]
\begin{center}
\includegraphics[trim=50mm 0mm 10mm 0mm,width=8cm, angle=270]{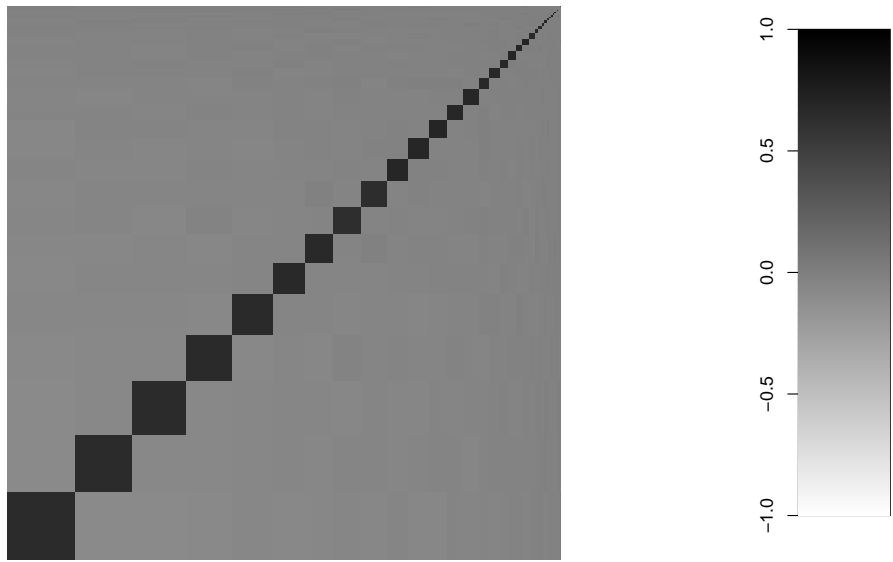}
\end{center}
\caption{\label{fig:bench}Matrix of correlations $\rho_{ij}$ for the communities set {\em a priori} in one of the instances of {\tt BENCH2} with $\mu=0.30$. We find that $82$ pairs $(C_i,C_j)$ outside the diagonal are not submodular ($\rho_{ij}>0$).}
\end{figure}

Figure~\ref{fig:bench} depicts the values of the correlation for all the pairs $(C_i,C_j)$ in one of the instances of {\tt BENCH2} with $\mu=0.30$. The partition that we considered here is the one set {\em a-priori} by the algorithm. We found 82 pairs of communities $(C_i,C_j), i \neq j$ that are not submodular (i.e., $\rho_{ij}>0$). The communities in these pairs will not be detected by modularity-based techniques, and this fact might explain why our fitness growth function can outperform them, when the real communities do not fulfill what we call the submodular condition. On the other hand, all the negative correlations are very close to zero, indicating that most of the pairwise unions would not produce a significant change in the modularity functional. This fact is in accordance with the observation in~\cite{clauset2010} that high-modularity partitions are prone to extreme degeneracy.

\begin{figure}[h]
\begin{center}
\includegraphics[width=16cm]{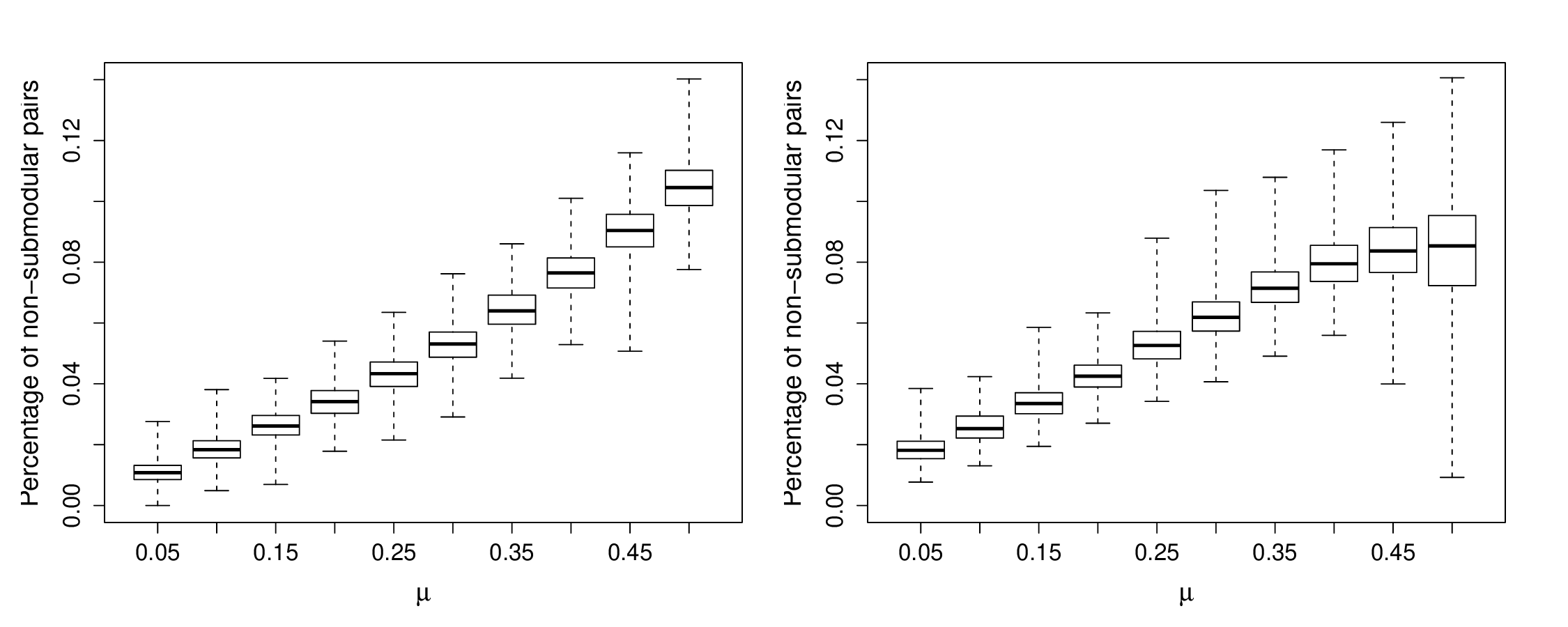}
\end{center}
\caption{\label{fig:submodulars}Boxplots representing the percentage of non-submodular community pairs $(C_i,C_j), i \neq j$ (where $\rho_{ij}>0$) for the 10,000 instances in {\tt BENCH2}, as a function of $\mu$. {\em (a)} Lancichinetti's {\em a priori} communities. {\em (b)} Communities obtained by our fitness growth process.

It is a remarkable fact that the original ({\em a priori}) communities are not submodular or, in other words, that the benchmark generates partitions for which modularity optimization techniques would tend to fail. We also point out that a similar plot for the partitions obtained by the Louvain algorithm would show a constant zero for the percentage of non-submodular pairs. This is a mandatory fact for any modularity maximization agglomerative technique which attains a local maximum.}
\end{figure}

In Figure~\ref{fig:submodulars} we analyze the existence of non-submodular communities for {\tt BENCH2}. The {\em y-axis} represents the percentage of not submodular pairs $(C_i,C_j), i \neq j$. For each $\mu$, the boxes represent the 1,000 network instances with that $\mu$. The left plot corresponds to Lancichinetti's {\em a priori} partition, while the right plot is for the communities that we obtain. The linear behavior of the percentage as a function of $\mu$ explains why modularity-based techniques tends to fail when the values of $\mu$ are bigger. In fact, in the Louvain algorithm the communities are merged until the condition $\rho_{ij} \leq 0$ is achieved.

\subsection{Robustness analysis}\label{sec:robustness}

In order to study the robustness of our method in real networks where
the actual communities are generally unknown, we propose to analyze the
mutual information between different partitions starting from randomly
chosen vertices, and observe the repeatability of the results.  The
studied networks include {\tt karate} club~\cite{zachary1977}, the
bottlenose {\tt dolphins} network~\cite{Lusseau_Newman_2004}, the
american college {\tt football} network in~\cite{girvan_2002}, an {\tt
e-mail} interchange network~\cite{guimera2003}, Erd\"os-R\'enyi
random graphs {\tt ER$*$}~\cite{ErdosRenyi59}, an instance from the {\tt BENCH3} benchmark with $\mu=0.40$ (see
section~\ref{sec:bench_desc}), a portion of {\tt
arXiv}~\cite{arxiv}, a collaboration network in Condensed Matter {\tt
ConMat}~\cite{girvan_2002}, and a portion of the World Wide Web network
{\tt WWW}~\cite{albert:tdofwww}.  Table~\ref{tab:real_N} shows the sizes of these
networks.

\begin{table}[h]
\small
\begin{center}
\begin{tabular}{l|c|c|c|c|c}
{\bf network} & $n$  & $m$ & $\langle|\mathcal{C}_{FGP}|\rangle$ & $stdev(|\mathcal{C}_{FGP}|)$ & $|\mathcal{C}_{Louvain}|$ \\ 
\hline \hline \hline
{\tt karate}   & 34 & 78 & 3.71 & 0.76 & 4 \\ 
{\tt dolphins} & 62 & 159 & 5.90 & 0.94 & 5 \\ 
{\tt football} & 115 & 613 & 10.19 & 1.20 & 10 \\ 
{\tt e-mail}   & 1133 & 5451 & 43.50 & 15.70 & 10 \\ 
{\tt BENCH3  } & 1024 & 5139 & 85.92 & 3.62 & 22 \\ 
{\tt arXiv}    & 9377 & 24107 & 1417.16 & 14.83 & 62 \\ 
{\tt CondMat}   & 36458 & 171736 & 4425.65 & 40.97 & 802 \\ 
{\tt WWW}      & 213715 & 446916 & 12655.29 & 28.35 & 358 \\ 
\hline
{\tt ER100}    & 100 & 508 & 11.97 & 3.39 & 8 \\
{\tt ER1k}     & 1000 & 5111 & 96.41 & 65.73 & 16 \\
{\tt ER10k}    & 10000 & 100261 & 919.24 & 800.46 & 10 \\
\end{tabular}
\caption{Summary of results for the analyzed networks. The columns represent: network size (number of vertices and edges), average number of communities found with the Fitness Growth Process and standard deviation, and the amount of modules discovered by Louvain's algorithm\label{tab:real_N}} 
\end{center}
\end{table} 

\begin{figure}[h]
\begin{center}
\includegraphics[width=12cm,angle=270]{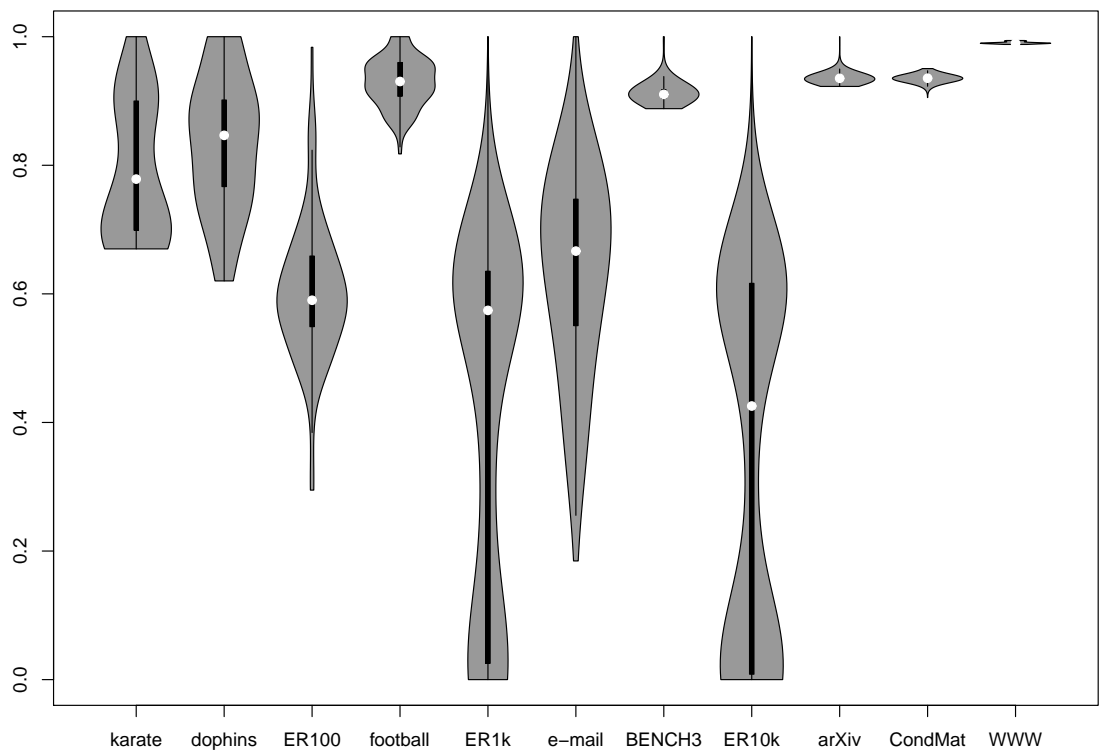}
\caption{Boxplots (with density) representing the results for different real networks and some Erd\"os-R\'enyi random graphs. The networks are spread over the {\em x}-axis. The boxplots and densities show the mutual information between the partitions obtained when starting from different vertices and a reference partition.\label{fig:robustness}}
\end{center}
\end{figure}

Figure~\ref{fig:robustness} shows the boxplots, together with the density
functions, of the mutual information for each network.  In each of them we
picked a random vertex, run the algorithm, and took the resulting
partition as the {\em reference partition}.  Then we started the
algorithm from other vertices, and measured the mutual information
between these partitions and the reference partition.  In small networks
we considered all the vertices, and just $1000$ different vertices for
{\tt arXiv} and {\tt ConMat} networks, and $48$ for the {\tt WWW}
network.
The fact that we just consider one reference partition to compare with
the others and do not make an all pairwise comparison is justified by
the transitivity relationship that we found
in~\ref{sec:mutual_inf_transitivity}.

The first observation of Figure~\ref{fig:robustness} is that the~\cite{ErdosRenyi59} random graphs ({\tt ER100}, {\tt ER1k}, {\tt
ER10k}) give a wide range of values of mutual information when the robustness analysis is
performed. This is an expected result, as it is in accordance with the fact that $ER$ graphs do not have a community structure, as~\cite{Lanci2011} points out. In fact, 
the amount of communities found is also very variable (see  Table~\ref{tab:real_N}), varying from $1$ to $1893$.

The {\tt e-mail} case is also remarkable because the mutual
information yields a wide range of values; this fact points out a
probably poor community structure in this network. The other networks
present high values of mutual information with small dispersions (i.e.,
boxplots are quite narrow). This trend is even more noticeable for the
large networks. In fact, the {\tt WWW} is an interesting case because
all the mutual information values that we found lay around its median
value of $0.989$ with extremes at  $0.989\pm0.02$, which means -by
transitivity- that the different partitions found when starting the
process from different vertices, are quite similar between them.

\subsection{Application to a collaboration network}

\begin{figure}[h]
\begin{center}
  \hfill
  \begin{minipage}[t]{140pt}

\includegraphics[width=5.0cm]{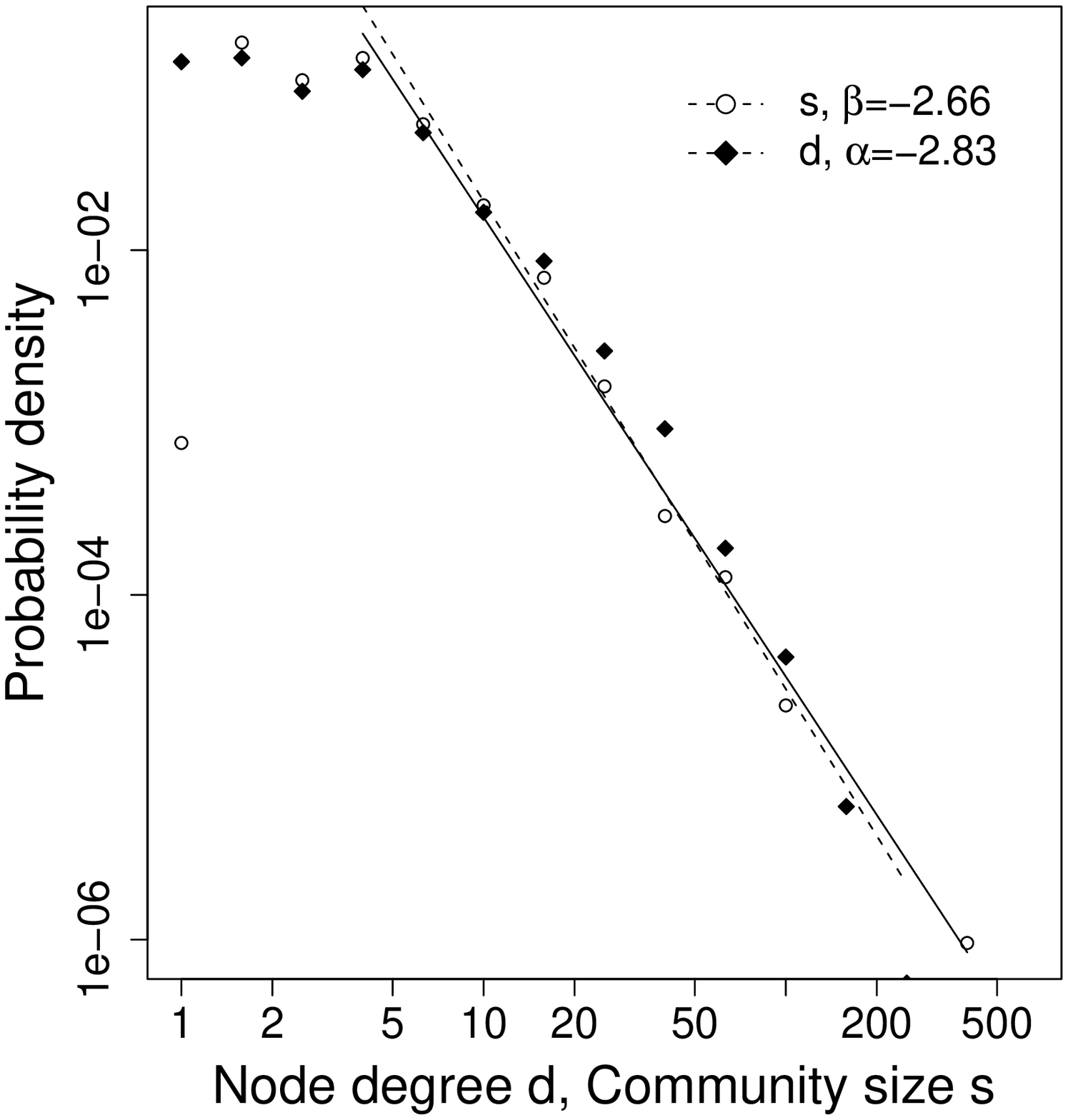}
  \end{minipage}
  \hfill
  \begin{minipage}[t]{230pt}
\includegraphics[height=5.0cm]{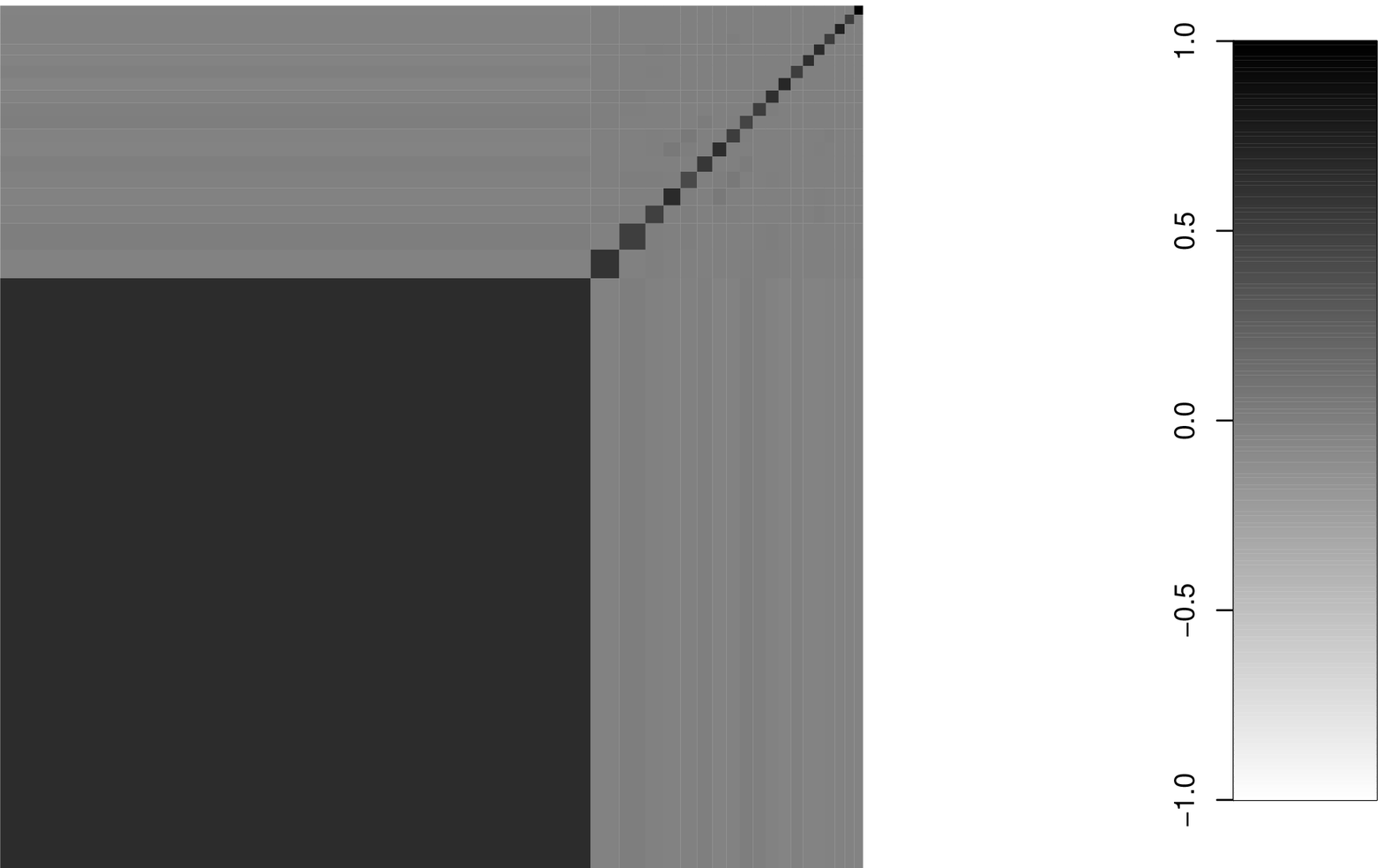}
  \end{minipage}
  \hfill
\caption{\label{fig:condmat_commsize_dist} (a) Community size and vertex degree distribution for the collaboration network CondMat. The histograms were built with a log-binning procedure. (b) Edges density between communities in terms of a correlation between Bernoulli variables, for the 20 biggest communities in CondMat.}
\end{center}
\end{figure}

Finally, we applied our algorithm to a network of coauthorships from the Condensed Matter E-Print Archive. We analyzed the giant component of the network, composed by 36,458 vertices and 171,736 edges. The result was a partition with 4425 communities, whose distribution follows a power-law on the community size (see Figure~\ref{fig:condmat_commsize_dist}.a) which may be due to the self-similarity of the network~\cite{havlin_song}. We remark the strong coincidence between the exponents on both distributions.

While the biggest community in this network contains about $31\%$ of the graph edges ($53880$ internal connections), it only has $406$ vertices (the $1.1\%$). Evidently, this community has a strong cohesion.

Figure~\ref{fig:condmat_commsize_dist}.b depicts the density of connections between all pairs of communities $C_i$ and $C_j$, in terms of the correlation $\rho_{ij}$ between two Bernoulli variables defined in~\ref{sec:corr_based}. The strong correlation in the diagonal implies a high density of edges inside the communities. The correlation values close to zero outside the diagonal imply a random amount of inter-community edges, similar to the expected amount in a null model graph.

\section{Conclusions}

The work by~\cite{Lanci2009} suggests the possibility of using different fitness functions for detecting local communities under a general procedure. In this work we have defined a fitness function $H_t$ and shown that it is essentially equivalent to the original one, which depends on a resolution parameter $\alpha$.
Then we proved an important fact: neither of the parameters (neither $\alpha$ nor $t$) play an important part in the vertex selection criterion, but only in the termination decision. This means, for example, that we can obtain a local community $C_{t}$ for some $t$, and then build the local community for $t'>t$ by taking $C_{t}$ and continuing the process until $t'$. So we proposed an unique fitness growth process which finds an ordering of the vertices such that the different communities lie one after the other. This sequence is the input of a three-staged algorithm that extracts a community partition of the graph. The algorithm is freely available to the scientific community as an open-source software which can be downloaded from {\tt http://code.google.com/p/commugp/}.

We also exploited a benchmark of heterogeneous graphs to test our method. On one side, we tested the correctness of the results by comparing them against communities defined {a priori}. On the other side, we gave an explanation on why global methods tend to fail on some heterogeneous networks. These ideas were illustrated by the use of a correlation measure and of normalized mutual information.

Finally we showed that the method is robust for many real networks. By analizyng random graphs, we pointed out that the behavior of the method may allow us to differentiate networks with a strong community structure from randomly connected ones.

As a future work we plan to study different ways of changing the vertex selection criteria of the growth processes, in order to avoid vertex eliminations. We also intend to extend the results for detecting situations of overlapping communities.

\ack This work was partially funded by an UBACyT 2010-2012 grant (20020090200119). M.G. Beir\'o acknowledges a Peruilh fellowship.

\section*{References}

\bibliographystyle{apalike}
\bibliography{comm}

\end{document}